\documentclass[slac_one]{revtex4}
\usepackage{graphicx}
\usepackage{fancyhdr}
\usepackage{amsmath}
\newcommand{\DM}{Dark Matter}
\newcommand{\AM}{antimatter}

\begin{document}

\title{Antimatter and \DM\ search in space with AMS-02} 

%

\author{Francesca R.~Spada (for the AMS-02 collaboration)}
\affiliation{Istituto Nazionale di Fisica Nucleare, Rome, Italy}
%

\begin{abstract}

AMS-02 is a space-borne magnetic spectrometer designed to measure with very high accuracy the composition of Cosmic Rays near Earth. With a large acceptance of 5000 cm$^2$, an intense magnetic field from a superconducting magnet (0.7 T) and a very efficient particle identification, AMS-02 will provide the highest precision in Cosmic Rays measurements up to the TeV region. 

During a three-years mission on board of the International Space Station, AMS-02 will achieve a sensitivity to the existence of anti-Helium nuclei in the Cosmic Rays of one part in a billion, as well as provide important information on the origin and nature of the \DM. 

We review the status of the construction of the AMS-02 experiment and its remarkable discovery capabilities.
\end{abstract}

\maketitle

\thispagestyle{fancy}

\section{THE STANDARD COSMOLOGICAL MODEL}
Recent precision measurements of cosmological quantities, such as the Cosmic Microwave Background temperature and polarization \cite{Spergel:2006hy, Page:2006hz, Netterfield:2001yq} and the Supernova luminosity-to-distance relationship, together with Large Scale Structure and structure formation studies \cite{Riess:2004nr, Perlmutter:1999jt}, allowed to build a \textit{standard cosmological model}, that describes a universe spatially flat, homogeneous and isotropic at large scales. 

\begin{figure*}[b]
\centering
\includegraphics[width=55mm]{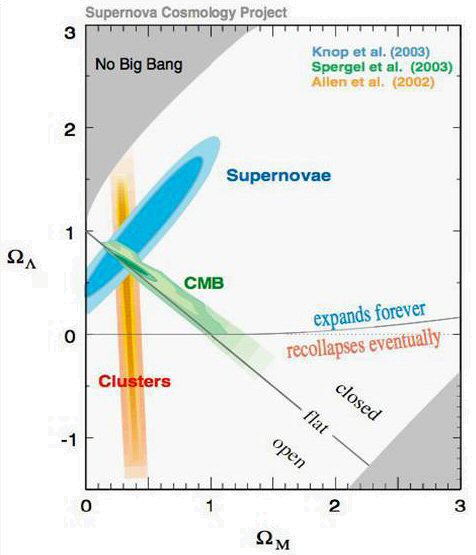}
\caption{Independent measurements of several cosmological quantities all point in the direction of a universe composed mostly of \DM\ and Dark Energy.} \label{fig:snproject}
\end{figure*}

Our universe consists of ordinary matter and radiation only for a 4.5\%: as shown in figure \ref{fig:snproject}, a considerable fraction of the total matter (23\%) is in the form of Cold \DM. The largest fraction (73\%) consists of Dark Energy, acting like a cosmological constant, whose origin is unknown \cite{Knop:2003iy}. The \AM\ content of the universe appears to be only 10$^{-6}$ of the matter content \cite{Alcaraz:2000ss}.

The aim of AMS-02 is to perform precision studies of Cosmic Ray properties such as composition, production, acceleration and propagation mechanisms.  Its large acceptance, energy resolution and particle identification capabilities allow AMS-02 to search for primordial \AM\ by direct detection of antinuclei, and for \DM\ annihilation products independently in different charged and neutral particles spectra.
\section{THE AMS-02 DETECTOR} 
AMS-02 is designed as a single particle spectrometer with large acceptance, high momentum range and efficient particle identification for reliable operation in space, to provide high statistics spectra of charged particles with nuclei and isotope separation, as well as gamma-ray measurements. It will carry out a 3-years mission taking data above the atmosphere on the International Space Station.

\begin{figure*}[t]
\centering
\includegraphics[width=50mm]{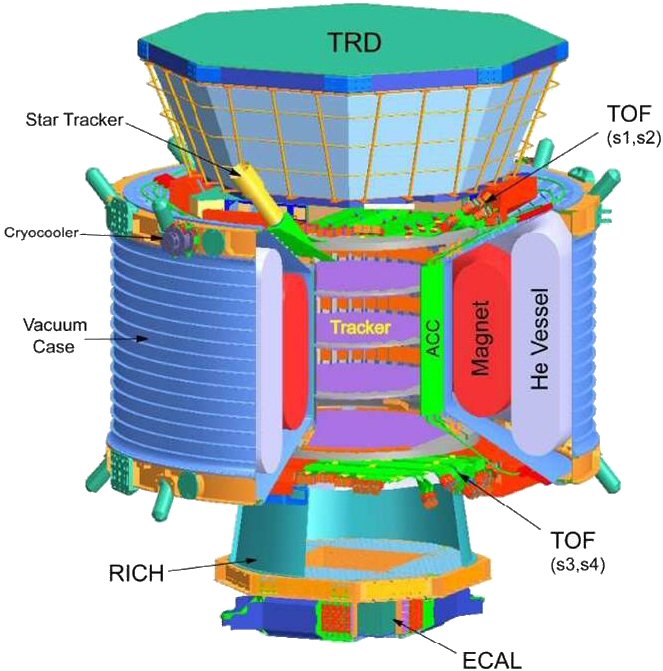}
\caption{The AMS-02 detector.} \label{fig:ams}
\end{figure*}

AMS-02 is shown in figure \ref{fig:ams}. It consists of a fully equipped tracker inside a superconducting magnet (the first ever to operate in space) providing a 0.8 T m$^2$ magnetic field. Rigidity is measured up to 3 TeV and nuclei and isotopes are separated up to 12 GeV/n for Z$\le$26 or A$\le$25 with a Ring Imaging aerogel/sodium-flouride Cherenkov detector (Agl/NaF RICH), combined with Time of Flight (TOF) and $dE/dX$ measurements. At 90\% positron efficiency, a proton suppression by 3-4 orders of magnitude is achieved with a lead/scintillating fibre sandwich 3D sampling calorimeter (ECAL) based on shower shape information and the matching of shower energy with track momentum. Further proton suppression by 2-3 orders of magnitude up to 300 GeV/c is achieved with a 20 layer fibre fleece Xe/CO2 proportional wire straw tube Transition Radiation Detector (TRD). The overall 10$^6$ proton rejection factor is essential for \DM\ search.

\section{PHYSICS GOALS}
\subsection{Indirect search for \DM}
A weakly interacting massive particle (WIMP) is the favoured \DM\ canditate, and supersymmetric extensions of the standard model such as mSUGRA contain a \textit{neutralino}  which is stable when R-parity is conserved. AMS-02 is capable of detectig WIMPs indirectly by measuring decay products of neutralino annihilation processes,
as an extra source for positrons, antiprotons, antideuterons and gammas on top of standard model predictions
from galactic propagation models. With annihilation cross sections proportional to the square of
the neutralino density, these measurements are also sensitive to \DM\ density fluctuations.

Recent measurements of gamma-ray fluxes \cite{Hunger:1997we} and the positron fraction \cite{Beatty:2004cy} have shown deviations from standard model expectations which are consistent with supersymmetric \DM, but controversial within the available statistics and without standard model explanation for the positron excess \cite{Strong:2004de, deBoer:2006tv}.

\begin{figure*}[t]
\centering
\includegraphics[width=45mm]{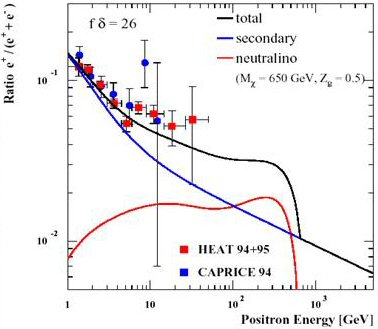} \hspace{7mm}
\includegraphics[width=46mm]{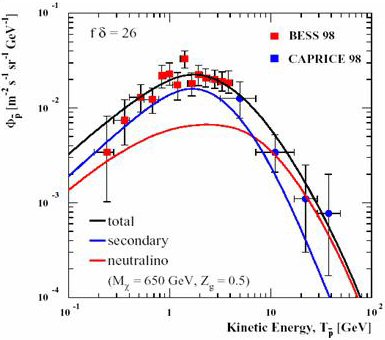}\hspace{7mm}
\includegraphics[width=40.5mm]{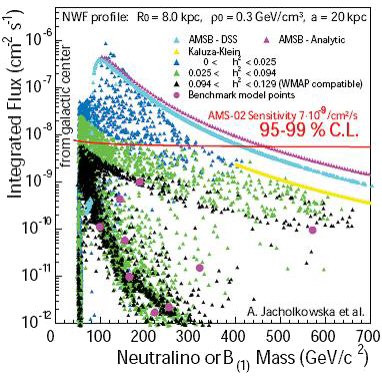}
\caption{Left: positron flux; center: antiproton flux, in the energy region accessible to AMS-02. The red solid line shows the expected signal for a 650 GeV neutralino. Right: part of the parameter space from new physics that si within reach of AMS-02 gamma flux measurements.}
\label{fig:dm}
\end{figure*}

AMS-02 will measure the positron spectrum with substantially improved precision for background and signal, in the energy region where an excess has been claimed. Moreover, the simultaneous measurement in other independent channels will increase the sensitivity to \DM\ detection. The spectra of positrons and antiprotons expected in the AMS-02 sensitivity region are shown in figure \ref{fig:dm}.

AMS-02 can also detect gamma rays up to 1 TeV either directly in the ECAL, or indirectly as an electron-positron pair after conversion in the material above the tracker, which amounts to 40\% of a radiation length. The part of the parameter space from new physics that is within reach of AMS-02 gamma measurements  in the case of cuspy halo profile or extra enhancements, and even for a standard NFW dark matter halo \cite{Jacholkowska:2005nz}, is also shown in figure  \ref{fig:dm}.

\subsection{Direct search for \AM}
\begin{figure*}[t]
\centering
\includegraphics[width=58mm]{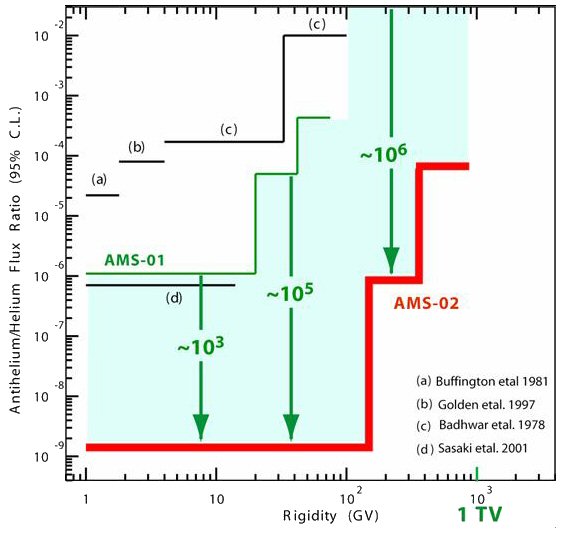}
\caption{AMS-02 sensitivity to the anti-Helium flux.} \label{fig:am}
\end{figure*}

In the Big Bang theory matter and antimatter are created with equal abundances, and the disappearance of antimatter requires barion number violation and another source of CP violation.
Antiparticles are indeed produced in collisions between high energy particles, and are observed in the Cosmic Rays. For example, 
\begin{align*}
&\phi(e^+) / \phi(e^-) \sim 10^{-1} \text{at 10 GeV} \\
&\phi(p) / \phi(\bar{p}) \sim 10^{-5} \text{at 10 GeV}.
\end{align*}
However, an anti-Helium nucleus has very low probability of being produced in collisions: 
\begin{align*}
\phi(\bar{He})/\phi(He)\sim  10^{-6} - 10^{-8}
\end{align*}
Anti-Helium detection in the Cosmic Rays would be a clear indication of the existence of an antimatter area somewhere in the universe.
AMS-02 will collect in three years $2 \times 10^9$ nuclei with energies up to 2 TeV, with a sensitivity that reaches anti-Iron. The limit put by the precursor flight AMS-01 will be increased of a factor $10^3$, meaning that, if no antinucleus is observed, there is no \AM\ to the edge of the observable universe (about 1000 Mpc). The limit on \AM\ presence they can be set by AMS-02 is shown in figure \ref{fig:am}.

\section{Perspectives}
AMS-02 will provide a coherent dataset of unprecedented precision for charged Cosmic Rays and gamma rays up to the TeV region. 

These data will confirm or disprove with high accuracy the excess in the HEAT positron data in the few GeV region.
The multichannel analysis will allow combined fits to the parameters of standard model extensions to establish the nature of \DM. Several models can be constrained and eventually ruled out.

AMS-02 will put a stringent limit on the presence of \AM: if no antinucleus is observed, the hypothesis of barion asymmetry will be strongly favoured as no antimatter areas are present in the observable universe.

In general, our knowledge of the Cosmic Rays physics will be improved. AMS-02 will perform an accurate study of composition (H, He, B/C, $^9$Be/$^{10}$Be) and energy spectra, and put significant constraints on galactic propagation model parameters.

It is also worth mentioning the possibility for a search of new types of matter (e.g. strangelets).

The detector integration will be completed in 2008, on schedule to be ready for launch with the Space Shuttle in 2010.


\end{document}